\documentclass[secnumarabic, graphics,floatfix, nofootinbib,tightenlines,nobibnotes, aps, prl, 12pt]{revtex4-1}
\usepackage{graphicx}
\usepackage[english]{babel}
\usepackage[utf8]{inputenc}
\usepackage{amsmath,amssymb}
\usepackage{amsfonts}
\usepackage{multirow}

\def\abs#1{\left|#1\right|}
\def\ave#1{\langle#1\rangle}

\begin{document}

\title{A derivation of the entropy-based relativistic smoothed particle hydrodynamics by variational principle}
\author{Philipe Mota$^{1}$}
\author{Weixian Chen$^{2}$}
\author{Wei-Liang Qian$^{3,2}$}

\affiliation{$^1$Centro Brasileiro de Pesquisas Fisicas, RJ, Brazil}
\affiliation{$^2$Instituto de F\'isica e Qu\'imica, Universidade Estadual Paulista J\'ulio de Mesquita Filho, SP, Brazil}
\affiliation{$^3$Escola de Engenharia de Lorena, Universidade de S\~ao Paulo, SP, Brazil}

\date{July 26, 2017}

\begin{abstract}
In this work, a second order smoothed particle hydrodynamics is derived for the study of relativistic heavy ion collisions.
The hydrodynamical equation of motion is formulated in terms of the variational principle.
In order to describe the fluid of high energy density but of low baryon density, the entropy is taken as the base quantity for the interpolation.
The smoothed particle hydrodynamics algorithm employed in this study is of the second order, which guarantees better particle consistency.
Furthermore, it is shown that the variational principle preserves the translational invariance of the system, and therefore improves the accuracy of the method.
A brief discussion on the potential implications of the model in heavy ion physics is also presented.

\end{abstract}


\maketitle
\newpage

\section{I. Introduction}

Hydrodynamics is one of the most venerable theoretical tools which has been playing an important role in our understanding of nature. 
Its applications are widely spread as well as deeply rooted in many distinct areas of physics.
For instance, the hydrodynamic description of heavy-ion nuclear collisions plays an essential part in the study of the properties of the hot and dense matter created at RHIC and LHC \cite{hydro-review-7,sph-review-1,hydro-review-6}, and it is further reinforced by the onging investigations of fluid/gravity duality \cite{adscft-fluidgravity-01,adscft-fluidgravity-02,adscft-fluidgravity-04,adscft-fluidgravity-06}.
Although the validity and the origin of the hydrodynamic model have been long under extensive discussions \cite{hydro-noneq-01,hydro-gradex-01,hydro-noneq-03}, simulation results \cite{sph-corr-1,hydro-vn-2,sph-corr-4,hydro-corr-2,sph-corr-ev-4} on azimuthal correlations for various systems have firmly demonstrated the success of the approach.
The smoothed particle hydrodynamics (SPH) \cite{sph-astro-1,sph-astro-2} is one of the oldest meshfree methods for the partial different equation which describes the dynamics of continuum media.
Distinct from any grid-based method such as the finite element method or the finite difference method, the SPH makes use of a set of arbitrarily distributed fluid elements, referred to as particles, to represent the system.
Each particle has a smoothing length, $h$, over which their properties are smoothed by a kernel function. 
In terms of the kernel function, the contribution of each particle is weighted according to their distance from the position in question.
Therefore, a physical quantity at a given spatial point is obtained by summing the relevant contribution from all the particles lying within the range of the kernel.
The SPH was firstly introduced to study astrophysical problems \cite{sph-astro-1,sph-astro-2,sph-astro-3}.
Nowadays, it is widely used to model fluid motion, as well as solid mechanics \cite{sph-app-1}.

Despite its wide applications, the original SPH suffers some inherent problems which lead to low numerical accuracy under certain circumstances.
Among others, particle consistency is one of the notable issues which reflects the discrepancy between the spatially discretized particles and the corresponding continuous form of the kernel function. 
Particle inconsistency demonstrates itself as the discretized SPH particles to be incapable of properly reproducing a constant function.
It usually results from the particle approximation process, which is closely associated with the boundary particles, non-uniformed particle distribution as well as the smoothing length.
The finite particle method (FPM) \cite{sph-algorithm-5,sph-algorithm-6} was proposed by Liu et. al. to improve the particle consistency.
The key idea of the approach is to perform the Taylor series expansion of the function to be approximated before multiplying both sides of the equality by the kernel function and integrating over relevant volume.
It was shown that the particle consistency is related to the order of the above Taylor series, and it is guaranteed independent of the specific form of the kernel function, neither to the particle distribution.

In implementing the SPH to the partial different equation, some rules are proposed to symmetrize or asymmetric the terms involving the gradient operator \cite{sph-algorithm-8}.
In the case of pressure gradient, the term is symmetrized in order to respect Newton's third law: the pair of forces acting on the two particles are equal in size but opposite in direction.
Alternatively, it is shown that the above result can be obtained naturally, if one derives the hydrodynamic equation by using the variational principle \cite{sph-algorithm-7,sph-vp-1,sph-1st}, which is a consequence that the system conserves linear and angular momentum.
For event by event fluctuating initial conditions, even though SPH particles are distributed uniformly at the initial instant, the distribution is likely to be disturbed as the system evolves in time.
Therefore, FPM formalism is particularly suitable to handle such physical system.
Since the momentum conservation is important for small systems created in the relativistic heavy ion collisions, one needs to develop a model which explicitly preserves the conservation law.
Owing to the complicated form of the FPM, it is not straightforward to guarantee the momentum conservation by symmetrizing certain physical quantities.
In order to apply the FPM to relativistic heavy ion collisions, one shall employ the variational principle to obtain the corresponding equation of motion.
In addition, the system created in the collision is of significantly high energy density with mostly vanishing baryon density, therefore the entropy should be chosen as the base of SPH algorithm.
This is the main goal of the present study.
In the following section, we briefly review the main feature of FPM and discuss its advantage. 
The entropy based hydrodynamical equation is derived section III by the variational principle.
Discussions and conclusions are given in the last section.

\section{II. The finite particle method}

For a physical quantity $f(x)$, the Taylor series expansion gives
\begin{eqnarray}
f(x) = \sum_{n=0} \frac{ (x-x_a)^n_i }{n!}(\partial^n_i f)_{x_a} .
\label{eq1}
\end{eqnarray}
Now we multiply both sides by a kernel function $W(x-x_a)$ and integrate over $x$ to obtain
\begin{eqnarray}
\int dx f(x) W(x-x_a) = \sum_{n=0} \frac{ (\partial^n_i f)_{x_a} }{n!} \int dx (x-x_a)^n_i W(x-x_a) .
\end{eqnarray}
If one retains the first term on the r.h.s. of the equality
\begin{eqnarray}
\int dx f(x) W(x-x_a) = f_a \int dx W(x-x_a) ,
\label{eq2}
\end{eqnarray}
and assumes that $\int dx W(x-x_a) = 1$, one restore the original SPH formula, namely,
\begin{eqnarray}
f_a = \int dx f(x) W(x-x_a) = \sum_b \frac{\nu_b f_b}{\rho_b} W(x_b-x_a) .
\end{eqnarray}
In the last step, one makes use of the particle approximation.

However, if one applies the following particle approximation directly to Eq.(\ref{eq2}),
\begin{eqnarray}
\int dx f(x) W(x-x_a) \rightarrow \sum_b \frac{\nu_b f_b}{\rho_b} W(x_b-x_a) \\
\int dx W(x-x_a) \rightarrow \sum_b \frac{\nu_b }{\rho_b} W(x_b-x_a) .
\end{eqnarray}
One obtains instead
\begin{eqnarray}
f_a = \frac{\sum_b \frac{\nu_b f_b}{\rho_b} W(x_b-x_a)}{\sum_b \frac{\nu_b}{\rho_b} W(x_b-x_a)} . \label{eq3}
\end{eqnarray}
It is noted that denominator on the r.h.s. of the equation is not exactly ``1" in practice, and Eq.(\ref{eq3}) is known as corrective smoothed particle method (CSPM) in literature which preserves the zeroth order kernel and particle consistency.

It is intuitive to generalize the above procedure to higher order. By retaining the r.h.s. of Eq.(\ref{eq1}) to the second order, one obtains
\begin{eqnarray}
\int dx f(x) W(x-x_a) = f_a \int dx W(x-x_a) + \partial_j f_a \int dx (x-x_a)_j W(x-x_a) . \label{philipe1}
\end{eqnarray}
If one replaces $W(x-x_a)$ by $\frac{(x-x_a)_i}{\abs{x-x_a}}W'(x-x_a)$ in the above equation, one has
\begin{eqnarray}
\int dx f(x) \frac{(x-x_a)_i}{\abs{x-x_a}}W'(x-x_a) = f_a \int dx \frac{(x-x_a)_i}{\abs{x-x_a}} W'(x-x_a) \nonumber \\ 
+ \partial_j f_a \int dx (x-x_a)_j \frac{(x-x_a)_i}{\abs{x-x_a}} W'(x-x_a) . \label{philipe2}
\end{eqnarray}
By implementing particle approximation, Eqs.(\ref{philipe1}-\ref{philipe2}) correspond to a matrix equation of $D+1$ dimension, with $D$ being the spatial dimension of the system, as follows
\begin{eqnarray}
\left[\begin{matrix}
\ave{f}_a
\\
\ave{f}_{a,j}
\end{matrix}\right]
= 
\left[\begin{matrix}
 \ave{1}_a & \ave{\Delta x_k}_a \\
 \ave{1}_{a,j} & \ave{\Delta x_k}_{a,j}
\end{matrix}\right]
\left[\begin{matrix}
f_a
\\
f_{a,k}
\end{matrix}\right] , \label{philipe}
\end{eqnarray}
where
\begin{eqnarray}
\ave{f}_a &\equiv& \sum_b \frac{\nu_b f_b}{\rho_b} W_{ab} ,\\
\ave{f}_{a,j} &\equiv& \sum_b \frac{\nu_b f_b}{\rho_b} \frac{(x_{ab})_j}{\abs{x_{ab}}}W'_{ab} .
\end{eqnarray}
For the Eq.(\ref{philipe2}), in general one may freely replace $W(x-x_a)$ by any basis function, and in particular, by $W'(x-x_a)$ as done in \cite{sph-algorithm-5}.
Our choice of $\frac{(x-x_a)_i}{\abs{x-x_a}}W'(x-x_a)$ garantees that $\frac{(x-x_a)_i}{\abs{x-x_a}}W'(x-x_a)$ is an even function as the kernel $W$.
The above equation can be used to express $f_a$ and $\partial_j f_a$ in terms of the properties of SPH particles,
\begin{eqnarray}
\left[\begin{matrix}
f_a
\\
f_{a,k}
\end{matrix}\right]
=
\left[\begin{matrix}
 \ave{1}_a & \ave{\Delta x_k}_a 
 \\
 \ave{1}_{a,j} & \ave{\Delta x_k}_{a,j}
\end{matrix}\right]^{-1}
\left[\begin{matrix}
\ave{f}_a
\\
\ave{f}_{a,j}
\end{matrix}\right] .
\end{eqnarray}
In one dimensional case, it gives
\begin{eqnarray}
f_a = \frac{ \ave{\Delta x}_{a,x} \ave{f}_a - \ave{\Delta x}_{a} \ave{f}_{a,x} }{ \ave{1}_a\ave{\Delta x}_{a,x} - \ave{1}_{a,x}\ave{\Delta x}_{a} } , \\
f_{a,x} = \frac{ \ave{1}_{a} \ave{f}_{a,x} - \ave{1}_{a,x} \ave{f}_{a} }{ \ave{1}_a\ave{\Delta x}_{a,x} - \ave{1}_{a,x}\ave{\Delta x}_{a} } .\label{Ph2D}
\end{eqnarray}
It is not difficult to see that for a constant function, the first line of Eq.(\ref{Ph2D}) naturally leads to the constant, while the second line guarantees a vanishing first order derivative.

In Fig.1 we show the SPH fit to the superposition of two random Gaussian functions by using standard SPH as well as FPM.
The upper panel corresponds to uniform particle distribution, and lower panel corresponds to non-uniform particle distribution.
\begin{figure}
\begin{tabular}{cc}
\begin{minipage}{160pt}
\centerline{\includegraphics[width=180pt]{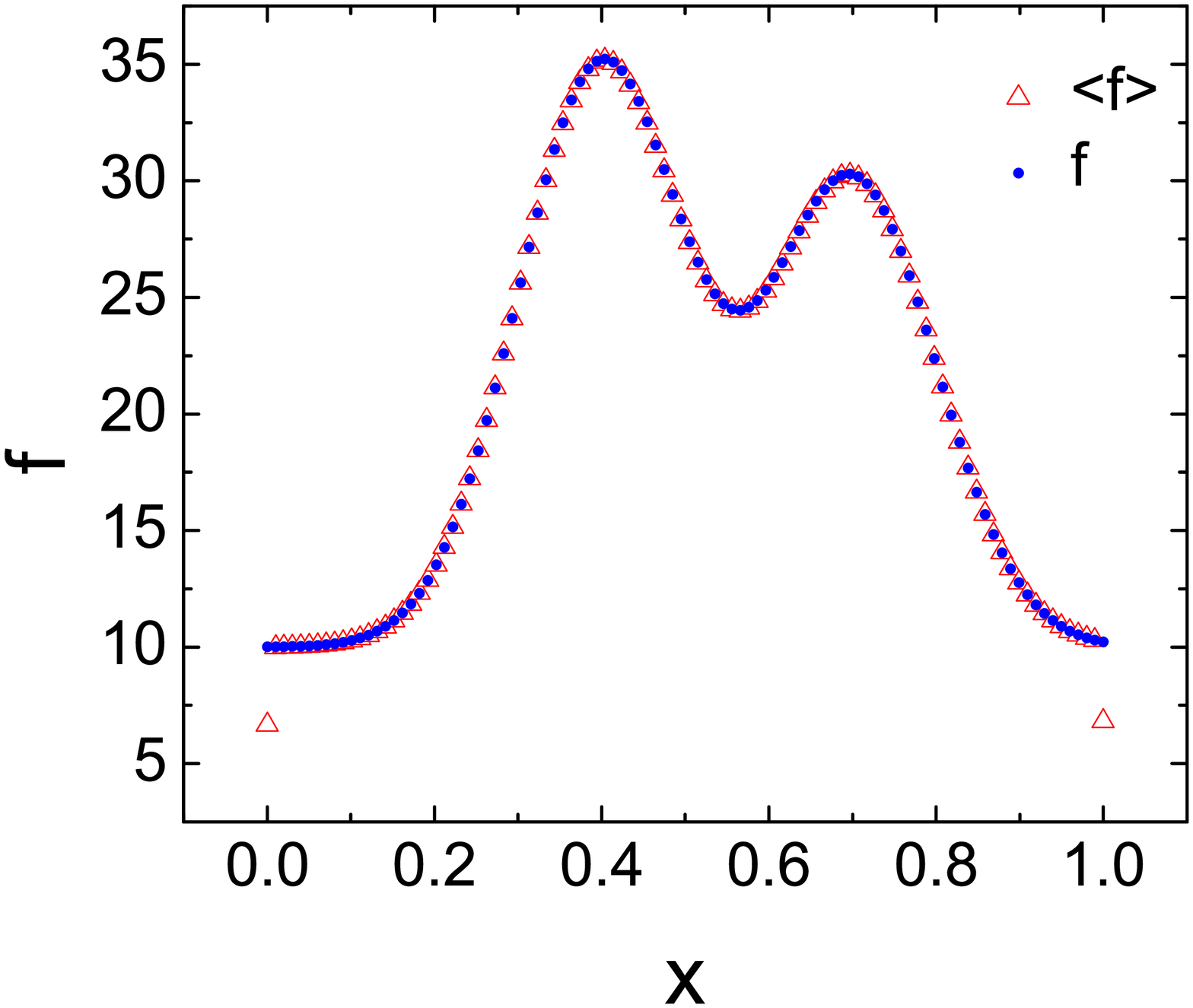}}
\end{minipage}
&
\begin{minipage}{160pt}
\centerline{\includegraphics[width=180pt]{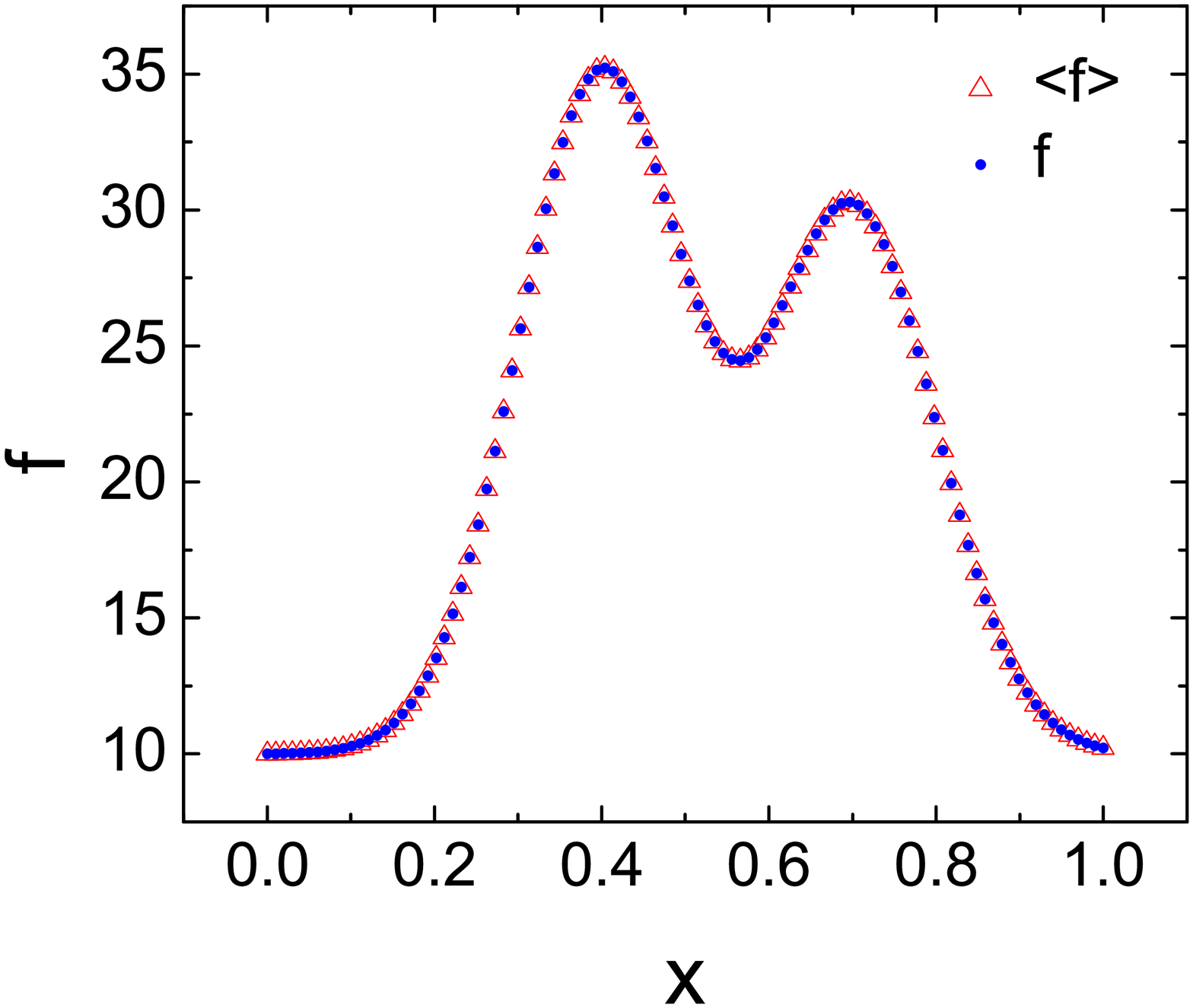}}
\end{minipage}
\\
\begin{minipage}{160pt}
\centerline{\includegraphics[width=180pt]{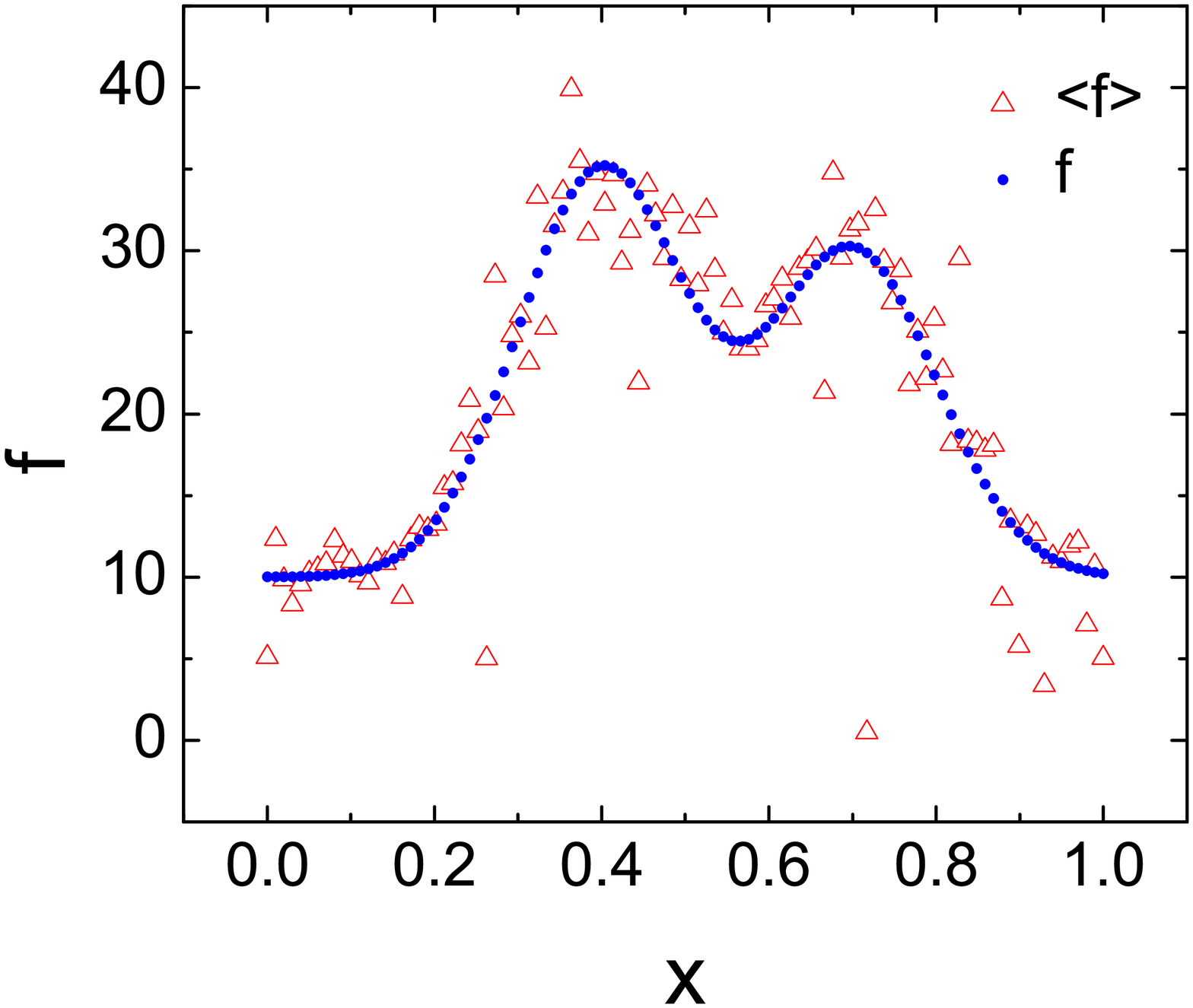}}
\end{minipage}
&
\begin{minipage}{160pt}
\centerline{\includegraphics[width=180pt]{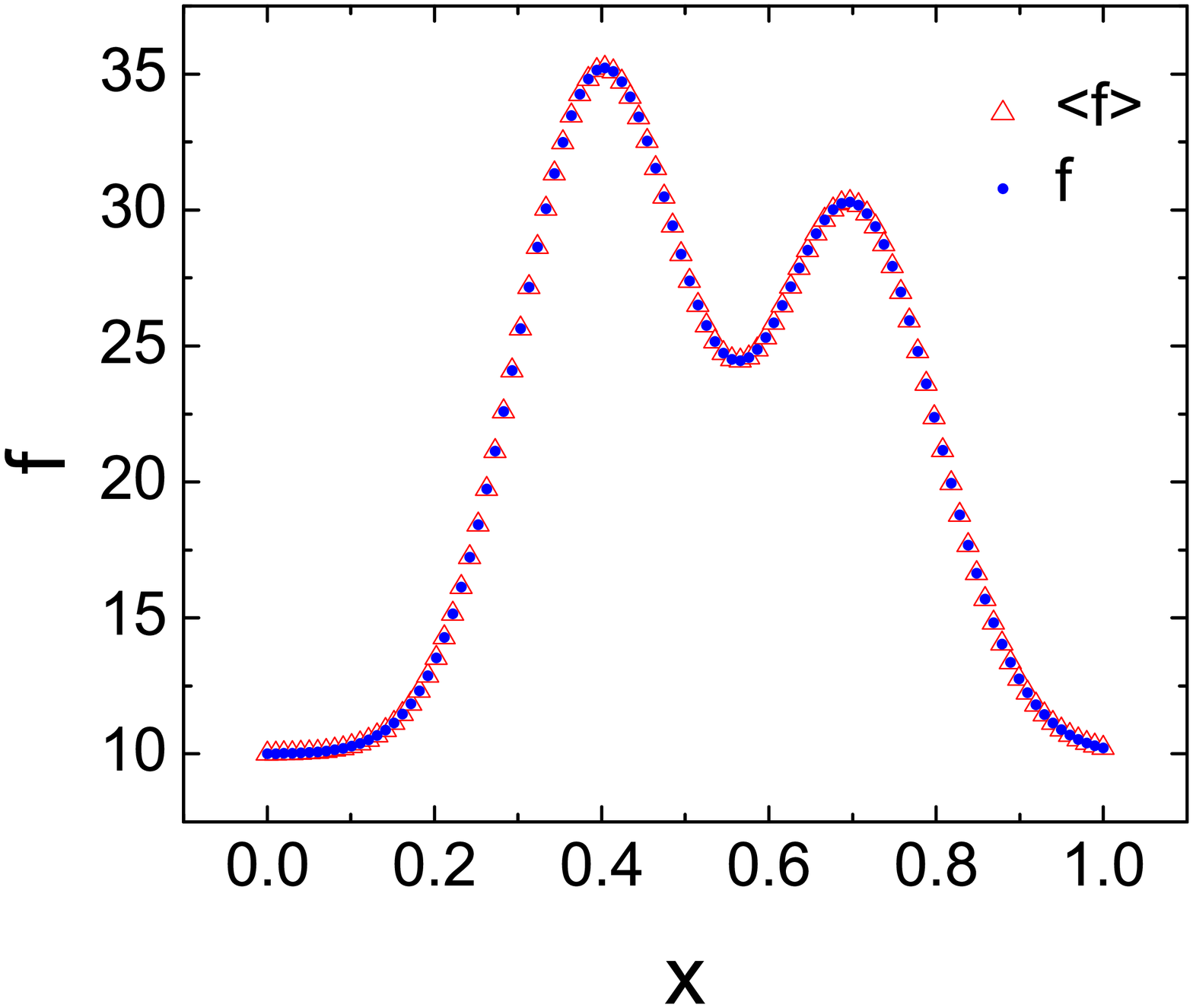}}
\end{minipage}
\end{tabular}
 \caption{(Color online) Plots for SPH fit to the superposition of two random Gaussian function by using standard SPH as well as FPM.
The original function is shown in blue dots, while SPH result is denoted by empty red triangles.
The left column is the results for standard SPH, and the right column is those for FPM.
The upper panel corresponds to uniform particle distribution, and lower panel corresponds to random particle distribution.
In the calculations, we make use of 1000 SPH particles with $h=0.002$ in all four cases.
}
 \label{fig1}
\end{figure}
One sees that even for uniform particle distribution, the standard SPH interpolation cannot properly reproduce the points on the boundary, as discussed in \cite{sph-algorithm-5}.
For non-uniform distribution, FPM is obviously superior to the standard SPH, which is the case that one frequently encounters in the heavy-ion nuclear collisions.
The above study is about fit to a given function, in what follows, we will explore the properties of the temporal evolution of the system.

\section{III. Hydrodynamic equation and temporal evolution}

The relativistic hydrodynamic equation for ideal fluid can be obtained by the conservation of energy-momentum flux \cite{sph-review-1},
\begin{eqnarray}
\frac{d}{d\tau}\left(\frac{(\epsilon+P)}{s}\gamma g_{ij}v^j\right)-\frac{1}{s\gamma}\partial_i P=0 . \label{rhydro}
\end{eqnarray}
where $\epsilon, P, s$ are the energy density, pressure and entropy density in the co-moving frame, $v^i, \gamma$ are the three velocity and gamma factor of the fluid element, $g_{ij}$ is the metric in Minkowski space.

It is noted that the conservation of the entropy flow is valid once there is no viscosity, and it is consistent with the standard form of entropy based SPH formula \cite{sph-review-1}, namely,
\begin{eqnarray}
s_i^* = \sum_j \nu_j W_({\bf r}_{\,i}-{\bf r}_{\,j};h) \, , \label{sstandard}
\end{eqnarray}
In the case of standard SPH, one substitutes the following symmetrized form for the pressure gradient \cite{sph-algorithm-8},
\begin{eqnarray}
(\partial {P})_i = \sum_j \nu_j s_i^* \left(\frac{{P}_i}{s_i^{*2}}+\frac{{P}_j}{s_j^{*2}}\right) \nabla_i W_({\bf r}_{\,i}-{\bf r}_{\,j};h)  ,\label{pgrad}
\end{eqnarray}
one leads to the following hydrodynamic equation in terms of degree of freedom of SPH particles.
\begin{eqnarray}
\frac{d}{dt}\left(\nu_i\frac{P_i+\varepsilon_i}   {s_i}\,\gamma_i\,{\bf v}_i\right)=\sum_j-\nu_i\nu_j \bigg[\frac{P_i}{{s_i^{\ast }}^{2}}+\frac{P_j}{{s_j^{\ast}}^2}\bigg]\, \nabla_i W({\bf r}_{\,i}-{\bf r}_{\,j};h)\, ,
\label{SPH_eq1}
\end{eqnarray}
where quantities with a superscript ``*" are evaluated in the laboratory frame and thus evaluated by using the SPH interpolation or by the equation of state (EoS), they are related to the corresponding quantity in the co-moving frame by a gamma factor (eg. $s_i^*=\gamma s_i$) due to Lorentz contraction.

Similarly, in the case of FPM, the pressure gradient on the r.h.s. of Eq.(\ref{rhydro}) can be written as
\begin{eqnarray}
(\partial {P})_i &=& P_{i,x} = \frac{ \ave{1}_{i} \ave{P}_{i,x} - \ave{1}_{i,x} \ave{P}_{i} }{ \ave{1}_i\ave{\Delta x}_{i,x} - \ave{1}_{i,x}\ave{\Delta x}_{i} } ,
\end{eqnarray}
where,
\begin{eqnarray}
\ave{P}_i &=& \sum_j \frac{\nu_j P_j}{\rho_j} W_{ij} ,\\
\ave{P}_{i,x} &=& \sum_j \frac{\nu_j P_j}{\rho_j} \frac{(x_{ij})}{\abs{x_{ij}}}W'_{ij} .
\end{eqnarray}

However, the above hydrodynamic equation does not take into consideration the momentum conservation.
Unlike the case of standard SPH, it is not obvious how to straightforwardly write down a symmetrized form as in Eq.(\ref{pgrad}) to guarantee that the resultant equation of motion respects the conservation law.
By looking closely at the r.h.s. of Eq.(\ref{SPH_eq1}), one observes that it can be written as
\begin{eqnarray}
\sum_j {\bf f}_{ij}\, .
\end{eqnarray}
with
\begin{eqnarray}
{\bf f}_{ij} = -\nu_i\nu_j \frac{P_i}{{s_i^{\ast }}^{2}}\, \nabla_i W({\bf r}_{\,i}-{\bf r}_{\,j};h)\, .\label{fold}
\end{eqnarray}
Since the kerner function $W$ is an even function, one finds
\begin{eqnarray}
{\bf f}_{ij} = -{\bf f}_{ji}\, .
\end{eqnarray}
In other words, the force excerted on $i-$th particle by $j-$th particle satisfies Newton's third law.
By using the variational principle, the translational variance of the Lagrangian density implies the momentum conservation or Newton's third law.
In what follows, we derive the hydrodynamic equation by using the variational approach.
Following \cite{sph-1st}, the action of the system can be written as 
\begin{eqnarray}
L_{SPH}(\{{\bf r}_i,{\bf {\dot{r}}}_i\}) = -\sum_i\left(\frac{E}{\gamma}\right)_i = -\sum_{i}\nu_{i}(\varepsilon/s^{\ast})_i \,, \label{L_SPH} 
\end{eqnarray}
where $E_i$ is the ``rest energy'' of the $i-$th particle \cite{classical-mechanics-goldstein}.
When applying the variational principle $\delta S_{SPH} = \delta \int dt L_{SPH}=0$, we note that one has $\delta E_i = - P_i \delta V_i$ in the co-moving frame, $V_i = \frac{\nu_i}{s_i}$ and $\delta \gamma = {\bf v} \cdot \delta {\bf v} \gamma^3$, which lead to
\begin{eqnarray}
0 = \delta S_{SPH} = -\int dt \left\{ \sum_i \delta {\bf r}_i \cdot \frac{d}{dt} \left[ \nu_i \left( \frac{P_i+\varepsilon_i}{s_i}\right)\gamma_i {\bf v}_i \right] + \sum_i \frac{\nu_i P_i}{({s_i}^*)^2} \delta  {s_i}^* \right\} \label{var1}.
\end{eqnarray}
If Eq.(\ref{sstandard}) were used, one would find
\begin{eqnarray}
\sum_i \frac{\nu_i P_i}{(s_i^*)^2} \delta s_i^* = \sum_{i,j} \left(\frac{\nu_i P_i}{(s_i^*)^2}\nu_j+\frac{\nu_j P_j}{(s_j^*)^2}\nu_i \right) \nabla_i W({\bf r}_{\,i}-{\bf r}_{\,j};h) \delta {\bf r}_i\, ,
\end{eqnarray}
and consequently Eq.(\ref{SPH_eq1}).

Now, to calculate the hydrodynamic equation for the FPM case in a one-dimensional system, we make use of Eq.(\ref{Ph2D}), namely,
\begin{eqnarray}
s_i^* &=& \frac{ \ave{\Delta x}_{i,x} \ave{s}_i - \ave{\Delta x}_{i} \ave{s}_{i,x} }{ \ave{1}_i\ave{\Delta x}_{i,x} - \ave{1}_{i,x}\ave{\Delta x}_{i} } , \label{sfpm}
\end{eqnarray}
on the r.h.s. of Eq.(\ref{var1}).
Before carrying out any explicit calculation, we note that in this case Newton's third law is guaranteed since Eq.(\ref{sfpm}) is translational invariant: it remains unchanged if all SPH particles shift the same amount $x_i \rightarrow x_i+X$.
To be specific, for any quantity $a_i = g_i \sum_j t_j W^{(e)}(x_{\,i}-x_{\,j};h)$ where $W^{(e)}(x_{\,i}-x_{\,j};h)$ is any even kernel function, it is straightforward to find
\begin{eqnarray}
\delta \left(\sum_i a_i\right) = \sum_{ij} (g_it_j+g_jt_i) {W^{(e)}}'(x_{\,i}-x_{\,j};h) \delta x_i \equiv \sum_j (f^{(e)}_{ij}+f^{(e)}_{ji}) \delta x_i \, .
\end{eqnarray}
Similarly, for any quantity $b_i = h_i \sum_j u_j W^{(o)}(x_{\,i}-x_{\,j};h)$ where $W^{(o)}(x_{\,i}-x_{\,j};h)$ is any odd kernel function, one has
\begin{eqnarray}
\delta \left(\sum_i b_i\right) = \sum_{ij} (h_iu_j-h_ju_i) {W^{(o)}}'(x_{\,i}-x_{\,j};h) \delta x_i \equiv \sum_j (f^{(o)}_{ij}+f^{(o)}_{ji}) \delta x_i \, .
\end{eqnarray}
In either case $f^{(e,o)}_{ij} = -f^{(e,o)}_{ji}$ is satisfied.
By a lengthy but straightforward calculation, one finds the hydrodynamic equation as follows
\begin{eqnarray}
\frac{d}{dt}\left(\nu_i\frac{P_i+\varepsilon_i}   {s_i}\,\gamma_i\,{v}_i\right)=\sum_j{f}^{(n)}_{ij} ,
\label{SPH_eq2}
\end{eqnarray}
where
\begin{eqnarray}
{f}^{(n)}_{ij} = -[(l^{(n)}_im^{(n)}_j+(-1)^{k^{(n)}}l^{(n)}_jm^{(n)}_i)]{W^{(n)}}'(x_{\,i}-x_{\,j};h) \, , \label{fnew}
\end{eqnarray}
with
\begin{eqnarray}
k^{(1,3,6,8)}_i &=& 1\nonumber\, , \\
k^{(2,4,5,7)}_i &=& 2\nonumber\, , \\
l^{(1)}_i &=& \frac{D_i\ave{s}_i}{B_i} \nonumber\, , \\
l^{(2)}_i &=& \frac{D_i\ave{\Delta x}_{i,x}}{B_i}  \nonumber\, , \\
l^{(3)}_i &=&  -\frac{D_i\ave{s}_{i,x}}{B_i} \nonumber\, , \\
l^{(4)}_i &=&  -\frac{D_i\ave{\Delta x}_i}{B_i} \nonumber\, , \\
l^{(5)}_i &=&  -\frac{C_iD_i\ave{\Delta x}_{i,x}}{B_i^2} \nonumber\, , \\
l^{(6)}_i &=&  -\frac{C_iD_i\ave{1}_i }{B_i^2} \nonumber\, , \\
l^{(7)}_i &=&  \frac{C_iD_i\ave{\Delta x}_i}{B_i^2} \nonumber\, , \\
l^{(8)}_i &=&  \frac{C_iD_i\ave{1}_{i,x}}{B_i^2} \nonumber\, , \\
m^{(1,3,5,6,7,8)}_i &=& \frac{\nu_i}{\rho_i}  \nonumber\, , \\
m^{(2,4)}_i &=& \nu_i  \nonumber\, , \\
W^{(1,6)}(x_{\,i}-x_{\,j};h) &=& \frac{x^2_{ij}}{|x_{ij}|}W'(x_{\,i}-x_{\,j};h)  \nonumber\, , \\
W^{(2,5)}(x_{\,i}-x_{\,j};h) &=& W(x_{\,i}-x_{\,j};h)  \nonumber\, , \\
W^{(3,8)}(x_{\,i}-x_{\,j};h) &=& x_{ij}W(x_{\,i}-x_{\,j};h)  \nonumber\, , \\
W^{(4,7)}(x_{\,i}-x_{\,j};h) &=& \frac{x_{ij}}{|x_{ij}|}W'(x_{\,i}-x_{\,j};h)  \nonumber\, ,\\
B_i &=& \ave{1}_i\ave{\Delta x}_{i,x} - \ave{1}_{i,x}\ave{\Delta x}_{i}\, ,\nonumber\\
C_i &=&  \ave{\Delta x}_{i,x} \ave{s}_i - \ave{\Delta x}_{i} \ave{s}_{i,x}\, ,\nonumber\\
D_i &=&  \frac{\nu_i P_i}{(s_i^*)^2}\, .
\end{eqnarray}

It is instructive to see how the resulting hydrodynamic equation reduces to the standard SPH formulae in its limit.
By comparing Eq.(\ref{fnew}) with Eq.(\ref{fold}), it is not difficult to find that Eq.(\ref{fold}) corresponds to the specific term ${f}^{(2)}_{ij}$ in one-dimensional case when one assumes $ \ave{1}_i \rightarrow 1$ and $ \ave{1}_{i,x} \rightarrow 0$. 
All other terms disappear when one makes use of the above limit as well as the symmetry of the kernel function, so that either $\ave{\Delta x}_{i} \rightarrow 0$ or $ \ave{\Delta x}_{i,x} \rightarrow 1$ takes place.

\section{IV. Concluding remarks}

To summarize, in this work, we make a preliminary attempt to study the implementation of FPM into the entropy-based SPH hydrodynamic model.
It is argued that the equation of motion obtained by using variational principle, though more complicated in its form, is more suitable to small and/or fluctuating systems where the conservation law plays a more stringent role in the dynamics.
We discussed possible implementations and in particular, derived the hydrodynamic equation of motion by using variational principle, where the momentum conservation of the system is assured.
It is shown how the obtained equation of motion reduces to the standard SPH form as a limit, which might be instructive to study the contributions of individual terms when one needs to introduce approximation for practical reasons.

Owing to the observation of the ``ridge" effect in two-particle correlation in relativistic heavy ion collision, the fluctuating initial conditions play an increasingly important role in the hydrodynamical description of nuclear collisions.
The AdS/CFT correspondence states the duality that two apparently distinct physical theories are closely connected.
It provides another insightful viewpoint of hydrodynamics as a gradient expansion in the long wavelength limit.
The applications of the SPH algorithm, from both aspects, strengthen the ongoing studies on heavy-ion physics as well as on gravitation theory.
The improvement of particle consistency brought by the FPM, therefore, can be significant in the context of precision and efficiency of the numerical approach.
It is interesting to implement the obtained equation of motion for realistic collision simulations, which will be carried out in our subsequent study.

\section*{Acknowledgments}
We are thankful for valuable discussions with Takeshi Kodama and Yogiro Hama.
We gratefully acknowledge the financial support from 
Funda\c{c}\~ao de Amparo \`a Pesquisa do Estado de S\~ao Paulo (FAPESP), 
Funda\c{c}\~ao de Amparo \`a Pesquisa do Estado de Minas Gerais (FAPEMIG), 
Funda\c{c}\~ao de Amparo \`a Pesquisa do Estado do Rio de Janeiro (FAPERJ),
Conselho Nacional de Desenvolvimento Cient\'{\i}fico e Tecnol\'ogico (CNPq),
and Coordena\c{c}\~ao de Aperfei\c{c}oamento de Pessoal de N\'ivel Superior (CAPES).

\bibliographystyle{h-physrev}
\bibliography{references_qian}{}

\end{document}